\documentclass[11pt,a4paper]{article}
\pdfoutput=1
\usepackage{jinstpub}
\usepackage{amssymb,amsmath}
\usepackage{multirow}
\usepackage{xcolor}
\usepackage{soul}
\usepackage{upgreek}

\title{Boron-10 conversion layer for ultra-cold neutron detection}

\author[1]{B.\,Clement}        
\author[1]{A.\,Bes}           
\author[1]{A.\,Lacoste}
\author[3]{R.\,Combe}
\author[2]{V.\,V.\,Nesvizhevsky}
\author[1]{G.\,Pignol} 
\author[1]{D.\,Rebreyend}
\author[1]{Y.\,Xi}

\affiliation[1]{LPSC, Universit\'e  Grenoble Alpes, CNRS/IN2P3, Grenoble, France}
\affiliation[2]{Institut Laue--Langevin, Grenoble, France}
\affiliation[3]{Normandie Univ, ENSICAEN, UNICAEN, CNRS/IN2P3, LPC Caen, Caen, France}

\emailAdd{bclement@lpsc.in2p3.fr}

\abstract{
We report on the development of a $^{10}$B conversion layer optimized for ultra-cold neutron detection with silicon detectors. The efficiency of this layer is high and roughly uniform over a large ultra-cold neutron velocity range.
The designed titanium-boron-nickel multilayer film was deposited on silicon using a microwave plasma-assisted co-sputtering method (first, for test purpose, on silicon wafers, then directly on the surface of a CCD sensor). The obtained sensor was then tested using both cold and ultra-cold neutrons. 
}


\begin{document}
%\notoc
\maketitle
\flushbottom
%\linenumbers

\section{Introduction}
\label{introduction}
When neutrons are slowed down to kinetic energies below $100$\,neV, 
they are totally reflected by most surfaces at any incidence angle. 
These \emph{ultra-cold neutrons} (UCN) constitute a sensitive tool to study fundamental interactions and symmetries \cite{Dubbers2011}. 
In particular, ultra-cold neutrons are used to study gravity in a quantum context \cite{Nesvizhevsky2002,Nesvizhevsky2010,Jenke2011,Pignol2015}.  
A neutron bouncing on top of an horizontal mirror realizes a quantum well problem and the vertical motion of the neutron bouncer has discrete energy states. 
The wave functions associated to the stationary quantum states have a spatial extension governed by the parameter $z_0 = (\hbar^2/2m_n^2g)^{1/3} \approx 6 \, \mu$m. 
Therefore, observing the spatial structure of the quantum states requires position-sensitive neutron detectors with a micrometric spatial resolution. 

The first generation of gravitational neutron quantum states experiments employed CR39 solid-state nuclear-track detectors. 
These are polymer plates used to record the tracks of heavy charged ionizing particles. 
The tracks are revealed after a chemical etching process and then read out with an optical microscope. 
To be sensitive to neutrons, the plastic plate is coated with a thin conversion layer converting the neutrons into energetic charged particles. 
Only four stable nuclei can convert slow neutrons to charged particles: $^3$He, $^6$Li, $^{10}$B and $^{235}$U. 
Isotopically enriched $^{235}$U was used in \cite{Nesvizhevsky2000,Baessler2011} whereas \cite{Jenke2013} used  $^{10}$B. 
The plastic detectors achieve a good spatial resolution ($\sigma \approx 2 \mu$m RMS in both cases) but suffer from significant deformation during the etching process. Such deformation can be corrected for, however this a complex procedure and a source of systematic uncertainties. Neutron detectors based on fined-grained nuclear emulsion can provide much higher resolution~\cite{Naganawa2018}. However, they lack the real-time readout capability. 

To overcome these severe limitations, semiconductor-based detectors have been recently investigated. 
In \cite{Jakubek2009,Jakubek2009b}, a silicon pixel device coated with $^6$LiF was exposed to ultra-cold neutrons. 
With a matrix of $256 \times 256$ square pixels with a pitch of $55 \, \mu$m, a spatial resolution of $5 \, \mu$m (FWHM) was measured. 
Sub-pixel resolution is achieved with the barycenter method of adjacent pixels. 
In \cite{Lauer2011}, a commercial CMOS chip with an active area of $3.2 \, {\rm mm} \times 4.8 \, {\rm mm}$ and a pitch of $3 \, \mu$m was tested. 
The conversion layer was coated on a AlMg$_3$ foil positioned in front of the sensor surface. 
Two different neutron converters, $^{10}$B and $^6$LiF were compared. 
Although the spatial resolution of such a device was not measured, it could be estimated to be $\sigma \approx 0.3 \, \mu$m. 
Last, \cite{Kawasaki2010} used a commercial back-illuminated CCD sensor ($512 \times 512$ pixels with a pitch of $24 \, \mu$m) 
covered with a neutron converter by vacuum evaporation. 
As converter material $^{10}$B and $^6$Li were compared and $^{10}$B was found to be more suitable. 
The spatial resolution was measured to be $\sigma = 3 \, \mu$m with a beam of cold neutrons. 

This article presents the development of a $^{10}$B-based layer to be used on position sensitive detectors to measure the wave functions of the bouncing neutrons in the GRANIT instrument~\cite{Roulier2015} at Institut Laue Langevin. 
Neutrons are converted in both a lithium and an helium nuclei in boron through the reaction:
\begin{equation*}
n+\textrm{$^{10}$B}\rightarrow \textrm{$^{7}$Li} + \textrm{$^{4}$He}~(+\gamma)
\end{equation*}
In 94\% of cases the lithium is produced in an exited state and an extra $477$\,keV gamma ray is emmited. The energies of the lithium nucleus and the alpha particle are $E_{Li}=841$\,keV and $E_\alpha=1471$\,keV, respectively. In the remaining 6\% of cases, no gamma ray is emitted and the particle energies are $E_{Li}=1014$~keV and $E_\alpha=1775$\,keV. 

The report will focus on the design, realisation and performance of the conversion layer. Spatial resolution will be discussed in a separate article later.
We chose a CCD sensor with a rectangular (rather than squared) active area, more adapted to the needed sensitive area of $300 \, {\rm mm} \times 0.5 \, {\rm mm}$. The CCD used is a Hamamatsu S11071-1106N sensor, without an entrance window. The measurement presented here were done using a specific electronics designed to read 8 sensors in series~\cite{Bourrion2018}.
To maximize the detection efficiency, without compromising on the spatial resolution, $^{10}$B should be used as a conversion material. 
Producing a stable and well controlled thin boron coating on silicon constitutes a major technological challenge. 
All previous studies used an evaporation technique. We have developed a reproducible process using plasma assisted physical vapor deposition. 

The article is organized as follows.
Section~\ref{design} presents the optimization of the conversion layer thickness and the calculation of the theoretical efficiency. Section~\ref{process} describes the new plasma coating process. Finally, section~\ref{neutrons} reports the first test of the prototype with both cold and ultra-cold neutrons
at ILL.
 
\section{$^{10}$B-based conversion layer design}
\label{design}
The design of the conversion layer takes into consideration several effects: the conversion layer must be sufficiently thick to absorb most of the neutrons but thin enough to let the secondary charged particles escape towards the silicon detector with sufficient energy.
The layer should not reflect neutrons either and remain chemically stable. As boron tends to get oxidized quickly, an external protective layer of a passivating material such as titanium or aluminum must be considered.
As the absoption cross-section goes as the inverse of the neutron velocity, the faster neutrons will tend to enter deeper into the boron layer and potentially escape undetected. 
One way to compensate for this effect is to add a thin UCN mirror between the boron layer and the silicon substrate. This layer could also improve bonding. The conversion layer will therefore consist of a multilayer of several materials on top of a silicon substrate.

\subsection{UCN interaction through thin layers}
Interaction of ultra cold neutrons with surfaces arise from their coherent scattering on nuclei. The surface of a material can be considered as a potential step. The interaction of a neutron with a thin layer is a simple one-dimensional well or barrier quantum problem. 
This potential, called optical Fermi potential, is a function of the scattering length $a$ and atomic density $\rho$:
\begin{equation}
V=\frac{2\pi\hbar^2}{m_n}\rho a,
\end{equation}
where $m_n$ is the neutron mass and $\hbar$ the reduced Planck constant.
In addition, absorption can be described by an imaginary potential, which will translate in an exponentially decreasing amplitude. This imaginary potential is defined in a similar way from the absorption scattering length $a_c$: 
\begin{equation}
W=-\frac{2\pi\hbar^2}{m_n}\rho a_c\textrm{ ,  } a_c=-\frac{\sigma k}{4\pi\hbar},
\end{equation}
where $\sigma$ is the absorption cross-section and $k$ the neutron momentum. Since the cross section varies as $1/k$, the product $\sigma k$ and therefore the potential is independent of the neutron velocity.
The total potential is then:
\begin{equation}
U=V-\imath W.
\end{equation}

Let's consider a structure composed of $n$ layers of material with thickness $\delta_{i=1..n}$ on a silicon substrate. This multilayer has $n+1$ interfaces that constitute as many quantum steps. 
The longitudinal momentum of a neutron with energy $E$ within the $i$-th layer is:
\begin{equation}
k_i^2 = \frac{2m_n}{\hbar^2}(E-V_i+\imath W_i),%%=\frac{2m_n}{\hbar^2}\sqrt{\left(E-V\right)^2+W^2}e^{\frac{i}{2}\arctan\left(\frac{W}{E-V}\right)}
\end{equation}
and the wave function is of the form:
\begin{equation}
\psi_i(x) = a_i e^{\imath k_ix} + b_i e^{-\imath k_ix}.
\end{equation}
Imposing continuity of the wave function and its derivative at each interface, one can compute the total reflection, transmission and absorption of the multilayer through the boundary matrix, assuming that the wave going out of the last interface has only a forward propagating part ($b_{n+1})$:
\begin{equation}
\left(
\begin{array}{c}t\\0\end{array}\right)
 = \prod_{i=n+1}^{1}\frac{1}{K_i+L_i}
\left(\begin{array}{cc}
K_i e^{-L_iG_i} &L_i e^{-K_iG_i}  \\
L_i e^{K_iG_i} &K_i e^{L_iG_i}  
\end{array}\right)
\left(\begin{array}{c}1\\r\end{array}\right)\equiv M\left(\begin{array}{c}1\\r\end{array}\right),
\end{equation}
with
\begin{equation}
K_i = k_i+k_{i-1}, \quad L_i = k_{i}-k_{i-1},\quad G_i = \sum_{j<i} \imath\delta_j,
\end{equation}
where $k_{0}$ and $k_{n+1}$ are the momenta in vacuum in front of the first layer and in the silicon substrate respectively and $r=\frac{b_0}{a_0}$, $t=\frac{a_{n+1}}{a_0}$.
The total reflection $R$, transmission $T$ and absorption $A$ of the multilayer are then:
\begin{equation}
\left\{\begin{array}{rcl}
R&=&|r|^2=\left|\frac{M_{21}}{M_{22}}\right|^2,\\
T&=&\left|\frac{k_0}{k_{n+1}}\right|\frac{1}{\left|M_{22}\right|} \textrm{ if } E-V_{n+1}>0,\,0\textrm{ if not,}\\
A &=& 1-R-T.
\end{array}\right.
\end{equation}

Quantum reflection of neutrons on the entrance layer might be a major source of inefficiency of the detector. Reflection happens both on the real and imaginary parts of the potential. For a negative real potential $U=V=-v$, $v>0$, which is the case of titanium, reflection is:
\begin{equation}
R_\textrm{real}=\frac{2E+v-2\sqrt{E^2+Ev}}{2E+v+2\sqrt{E^2+Ev}},
\end{equation}
whereas for an imaginary potential $U=-\imath w$:
\begin{equation}
R_\textrm{imag}=\frac{E+\sqrt{E^2+w^2}-2\sqrt[4]{E^4+E^2w^2}\cos\left(\frac{1}{2}\arctan\frac{w}{E}\right)}{E+\sqrt{E^2+w^2}+2\sqrt[4]{E^4+E^2w^2}\cos\left(\frac{1}{2}\arctan\frac{w}{E}\right)}.
\end{equation}
For a same potential absolute value, reflection on an imaginary potential is always larger than reflection on a negative real potential.

\subsection{Titanium-Boron-Nickel multilayer}
To satify the requirements of the conversion layer, the materials for the layers must be chosen carefully.
We used the model described in the previous section to optimize the width of the components of a $^{10}$B-based multilayer.
The Fermi potentials used in this document are summarized in table~\ref{tab-pot}.
\begin{table}[t]
\begin{center}
\caption[]{Effective potential for materials used in this study } 
\begin{tabular}{|l||cc|} \hline
Material & $V$ (neV) &  $W$ (neV)\\
\hline\hline
Titanium & -50.1 & 0.025  \\
Aluminum &  54.0 & 0.001  \\
Nickel   &  245.2& 0.030  \\
Silicon  & 54.1  & <0.001 \\
${}^{10}$B, 96\,\% enriched & 2.5 & 36.3   \\
\hline
\end{tabular}
\label{tab-pot}
\end{center}
\end{table}

\begin{figure}[t]
\begin{center}
\includegraphics[width=0.32\textwidth]{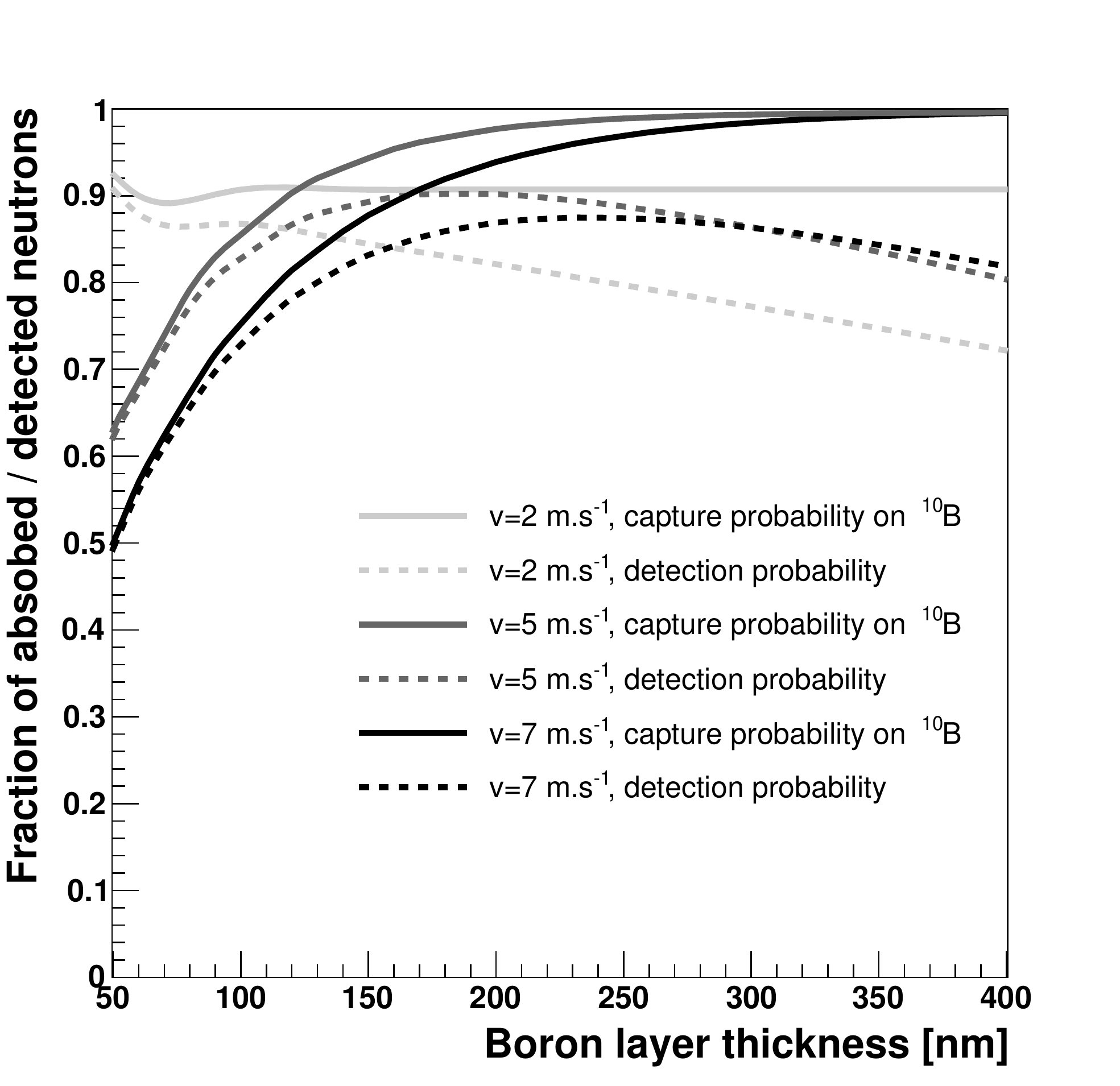}
\includegraphics[width=0.32\textwidth]{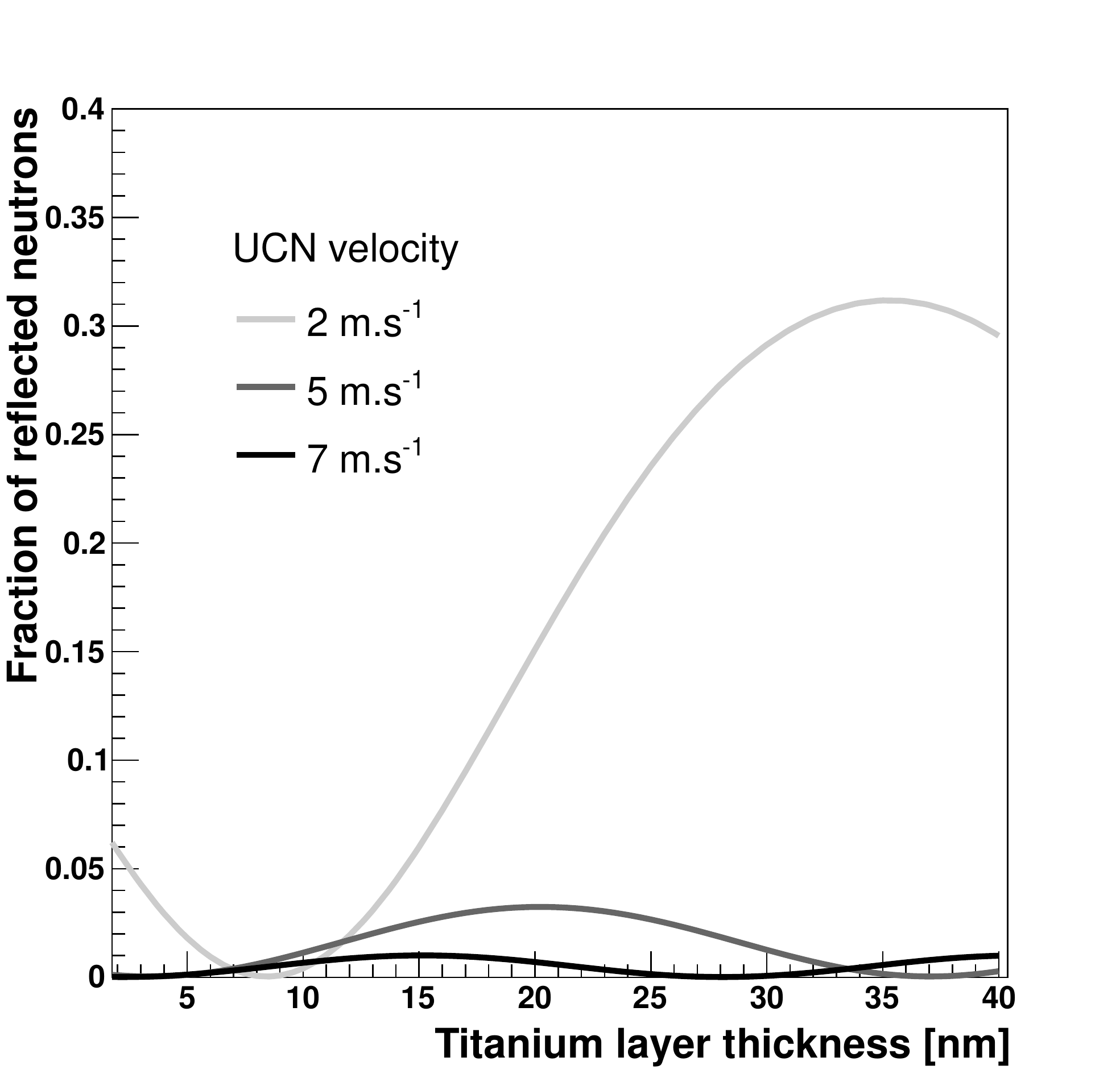}
\includegraphics[width=0.32\textwidth]{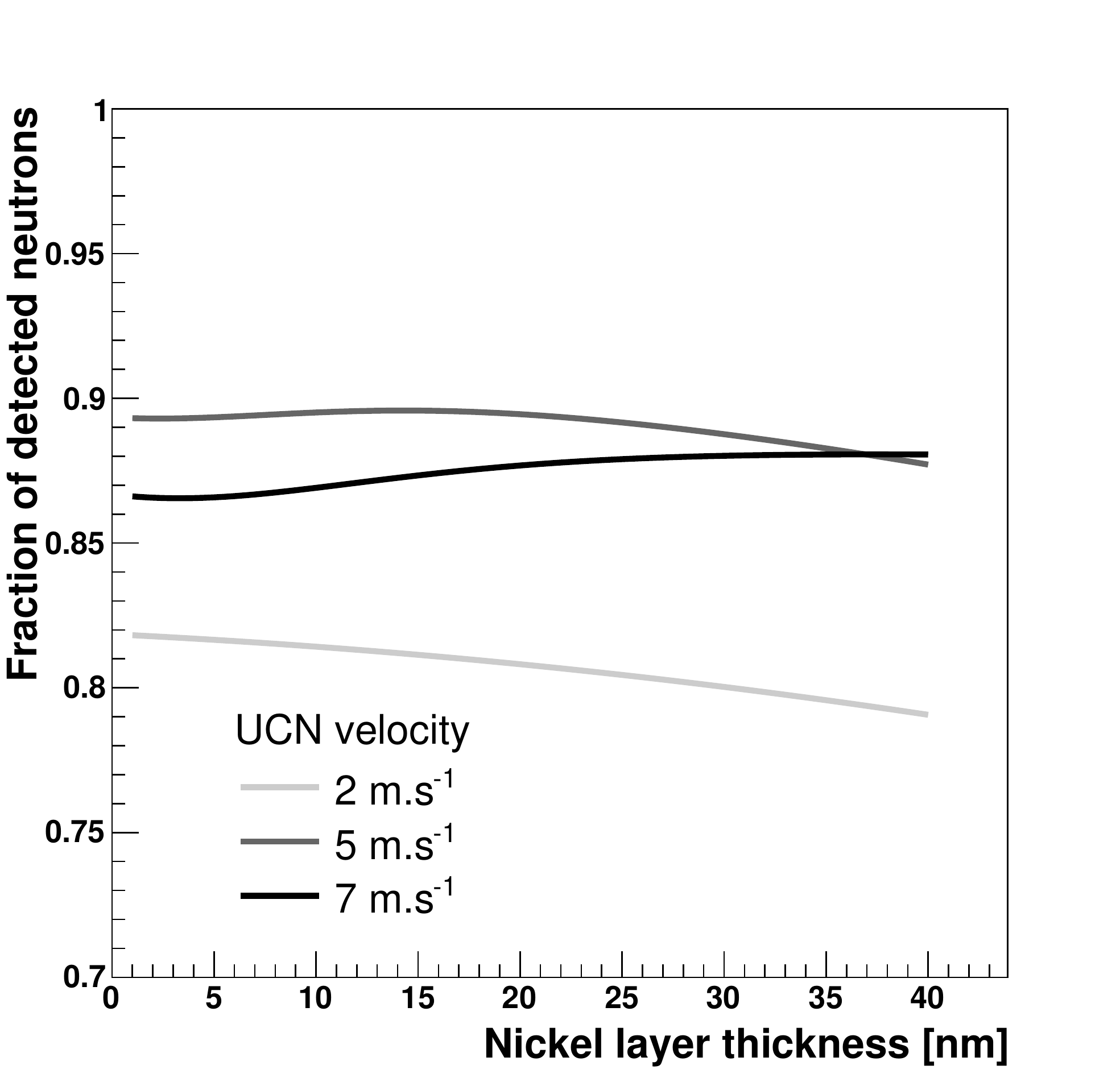}
\caption[]{Effects of the various layers on UCN as a function of the layer thickness: absorbtion of UCN and detection efficiency for pure $^{10}$B (left), reflection on the titanium layer on top of an infinite $^{10}$B layer (center) and the effect of a nickel layer on detection efficiency after a 200\,nm $^{10}$B layer (right).}
\label{fig-optim}
\end{center}
\end{figure}

The main optimisation will come from the thickness of the boron layer. The thicker the layer, the higher the conversion efficiency. However, the charged particles resulting from the conversion must reach the silicon sustrate with sufficient energy to be detectable. The detection efficiency if the probability for a neutron to be captured on boron and produce a charged particle able to reach th silicon with an energy above the threshold.
The efficiency should remain large in a range of UCN velocities up to 7\,m.s$^{-1}$ (the velocity cut-off for the GRANIT UCN source). Monte-Carlo simulation based on SRIM~\cite{SRIM} as well as NIST databases~\cite{NISTData} were used to estimate the energy loss. We fixed the energy threshold to 300\,keV (corresponding to what will be obtained experimentally later on). The result of this optimisation is shown in figure~\ref{fig-optim} (left) for both the capture probability (straight line) and the detection efficiency (dashed line). The optimum depends on the UCN velocity as the capture cross-section goes as its inverse. For a fixed boron thickness, slower neutrons interact on average closer to the entrance surface and the resulting charged particle have more material to cross resulting in higher energy loss.
The average velocity for the GRANIT source is around 5.1~m.s$^{-1}$ corresponding to an optimum of 180~nm but going up to 7~m.s$^{-1}$ corresponding to 230~nm.

A pure boron layer would not be practical. In our first trial, a 60~nm boron layer on silicon oxidized over a few weeks changing its color from blue to yellow. It is therfore important to add a protective layer on top of the boron.
Titanium was preferred to aluminum for the passivation layer as it has a large negative real potential and small absorption. If kept in a environment devoid of humidity, we can assume the titanium oxide passivation layer to be less than 2\,nm~\cite{Lee91} and have no significant impact on the UCN propagation.

As $^{10}$B has a real Fermi potential close to zero, but a large imaginary potential which is obviously a requirement for neutron conversion, it appears that a thin enough titanium layer could reduce the quantum reflection on the imaginary potential  as shown in figure~\ref{fig-optim} (center).
The optimal would be a layer of around $10$~nm but thicknesses up to $25$~nm show improvement.

Finally we consider the oportunity to add an intermediate layer between the boron and the silicon. The layer could have an effect on the boron adherance but could also be used to recover some neutrons by reflecting them on this layer. The reflective layer needs to have a large Fermi potential and good mechanical properties. Nickel is an obvious choice as it is commonly used for neutron guides.
Here again the thickness of the layer should not be exessive to limit energy loss of alpha and lithium ions, especially for slower neutrons, and the final detection efficiency is the pertinent parameter to optimize. The efficiency as a function of the nickel layer thickness for a 200\,nm enriched boron layer are shown in figure~\ref{fig-optim} (right). It is difficult to determine an optimum as the effect is strongly velocity dependent. For velocities lower than 4\,m.s$^{-1}$ no layer is required whereas 25\,nm to 35\,nm could help retrieve a few percent efficiency at 7\,m.s$^{-1}$.

\begin{figure}[t]
\begin{center}
\includegraphics[width=0.95\textwidth]{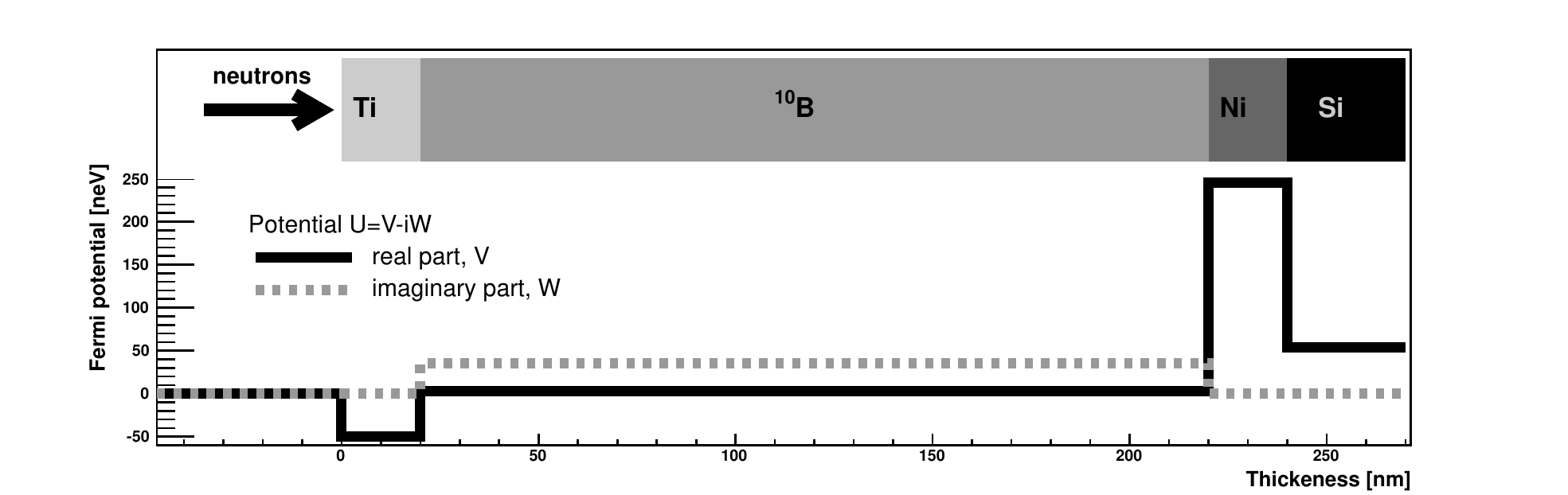}
\caption[]{Structure of the Ti/$^{10}$B/Ni mutlilayer in term of real and imaginary potential.}
\label{fig-potentials}
\end{center}
\end{figure}

The multilayer we effectivly realised is not the optimal one, as it has to include the limitation of the deposition process. Thicknesses below 20\,nm are difficult to ensure with a good uniformity. We therefore choose to use 20\,nm for both titanium and nickel layer. In both cases we chose the thinner we could safely realise short of removing the layer. 
For the boron layer, the UCN velocity spectrum of the GRANIT source, centered at 5.1\,m.s$^{-1}$~\cite{Roulier2015} would point to an optimal thickness of $180$\,nm. Nevertheless, as we might want to use this detector on slightly harder UCN spectrum at another facility, we choose a compromise of 200\,nm enriched boron.
Similarly the 20\,nm nickel layer will only affect marginally the slower neutrons and allow a small efficiency gain at higher velocities.
The potential step structure of such a multilayer is given in figure~\ref{fig-potentials} (right).

\begin{figure}[t]
\begin{center}
\includegraphics[width=0.48\textwidth]{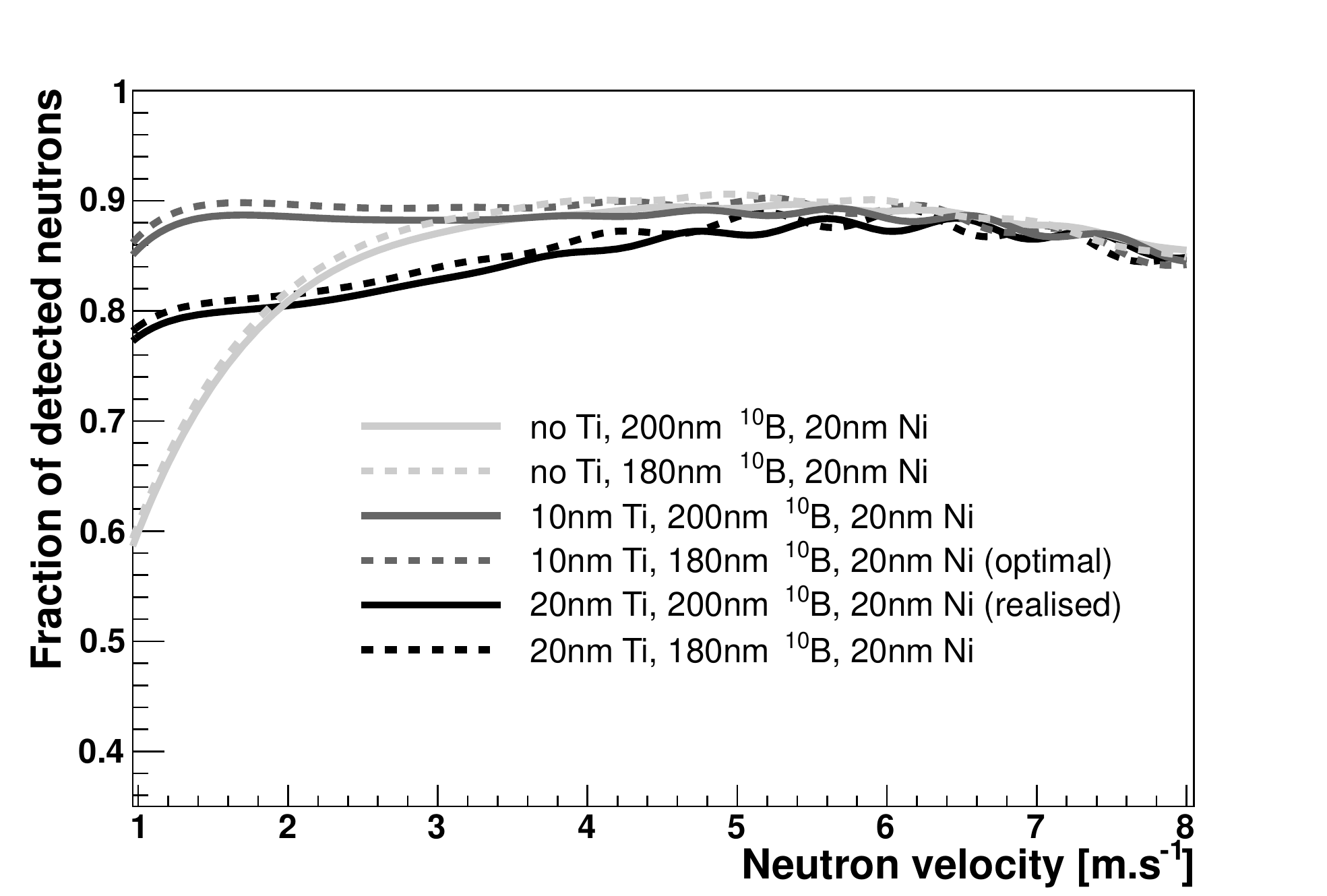}
\includegraphics[width=0.48\textwidth]{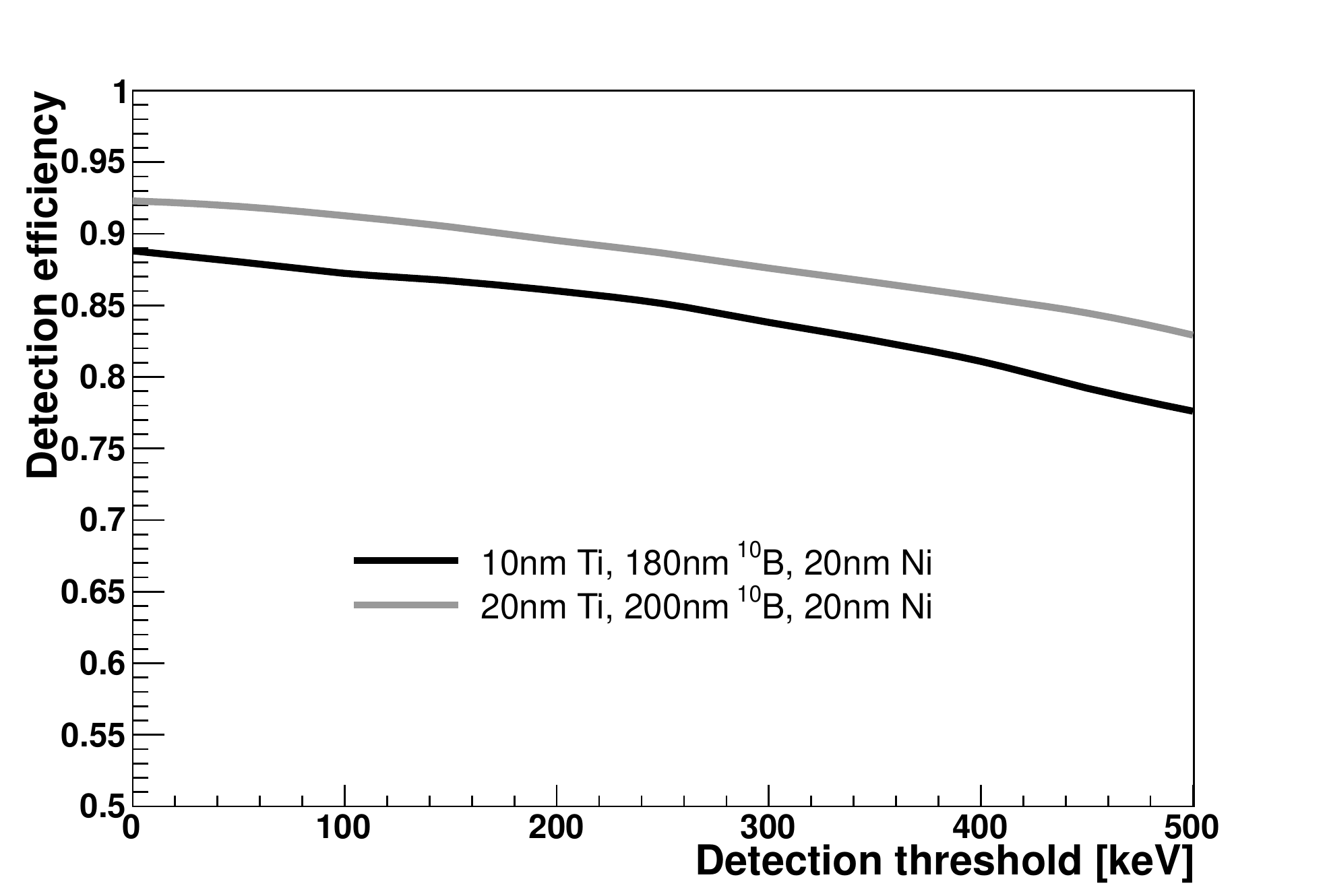}
\caption[]{Left: total efficiency as a function of the neutron velocity for various combinations of thiknesses of titanium and enriched boron. Efficiency includes neutron absorbtion probability and energy losses of secondary particles, for a detection energy threshold of 300\,keV.
Right: Detection efficiency as the function of the reconstruction energy threshold.}
\label{fig-efficiency}
\end{center}
\end{figure}

The final expected efficiency of the realised multilayer 
(20\,nm Ti, 200\,nm ${}^{10}$B, 20\,nm Ni) 
and of the optimal one 
(10\,nm Ti, 180\,nm ${}^{10}$B, 20\,nm Ni) 
are given in figure~\ref{fig-efficiency} (left).
Note that for defining this optimal, the nickel thickness was kept at the compromise value of $20$\,nm and not studied further as the effect remains, in any event, marginal. Figure~\ref{fig-efficiency} (right) present the depedence of the efficiency on the energy thresold, which itself will strongly depend on the nature of the detector and its associated electronics.

Averaging over the velocity spectrum for the GRANIT UCN source (gaussian distribution centered at 5.1\,m.s$^{-1}$ with a standard deviation of 1.6\,m.s$^{-1}$), the final efficiency of the realised multilayer  is expected to be $86\pm 1\%$. This error assumes a $20\%$ error on the spectrum parameters.
For the optimal multilayer, the average efficiency would reach $89\pm 1\%$.
The small error results from the low dependency of the efficiency on the velocity in the pertinent range.

\section{Boron coating with plasma assisted PVD}
\label{process}
Thin films were deposited on room temperature substrate by microwave plasma-assisted co-sputtering method. 
The principal feature of this technology described in greater detail in~\cite{Lequoc2011,Lequoc2012} is the separation of the plasma production (regulated by the microwave power) 
from the polarization of the surfaces involved in the deposition process. This separation allows, on the one hand, an independent and individual polarization of a set 
of targets making possible the individual control of their sputtering rate through the potential which is applied to each of them. On the other hand, 
the substrate can be either polarized or kept at ground potential according to the desired ion assistance during plasma processing. 
Moreover, the decoupling of the process parameters enables the implementation of in-situ sequential processes such as the substrate cleaning, 
targets cleaning and multilayer deposition without any exposure of samples outside the plasma reactor. 

\begin{figure}[t]
\begin{center}

\includegraphics[width=0.65\textwidth]{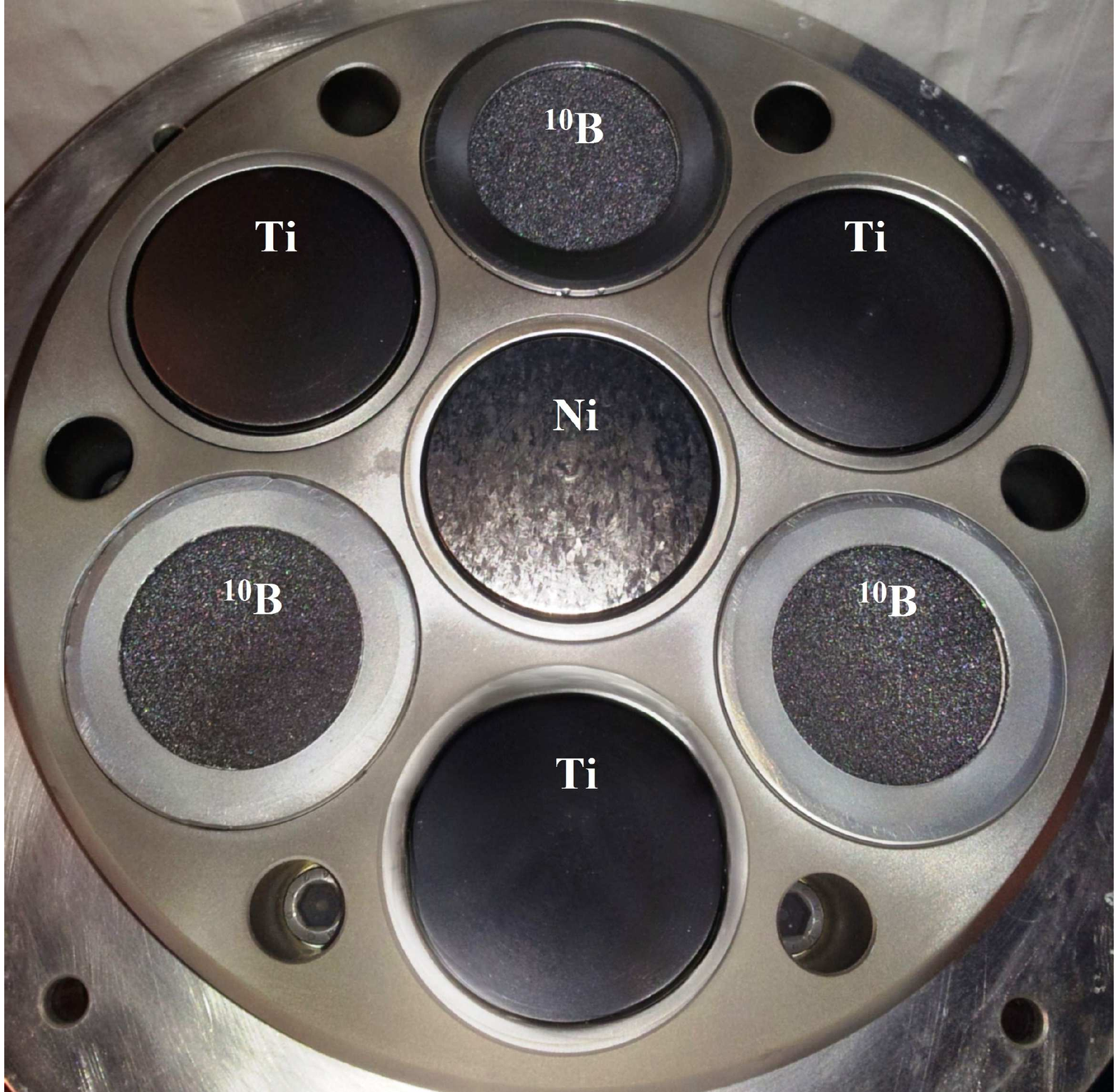}
\caption[]{Targets holder for multilayer film (Ni/$^{10}$B/Ti) deposition.}
\label{fig-targets}
\end{center}
\end{figure}

Multilayer films Ni/$^{10}$B/Ti were directly deposited onto CCD sensors by successive sputtering of Ni, $^{10}$B and Ti targets. 
For all process steps, an Ar plasma was sustained by 2\,kW of microwave power through 24 wave couplers evenly distributed on the plasma chambre wall. Vacuum and working pressures were of $10^{-4}$\,Pa and 0.13\,Pa, respectively. 
The multi-targets holder (see figure~\ref{fig-targets}) is placed on top of the chamber at an adjustable distance away from the substrate on which the films are ultimately grown. 
After a prior cleaning process of 10\,min in an ethanol sonic bath, the substrate is submitted during 20\,min in Ar plasma to an ion bombardment with an energy around 50\,eV. 
This results from 30\,V of DC self-bias (by applying 50\,W of RF power) added to 20 V of plasma potential.

During the deposition process, one Ni target, three $^{10}$B and three Ti targets, with a 99.99\% purity for each of them, are positioned 
on the target holder in the configuration shown in figure~\ref{fig-targets}. They were consecutively biased at different voltages and for different 
times: $-100$\,V (DC bias) during 23\,min for Ni target, $-440$\,V (RF bias) during 240\,min for $^{10}$B and $-100$\,V (DC bias)/18\,min for Ti. 
For a distance between the targets and substrate fixed at 150\,mm, the above process parameters were determined by a prior calibration 
operation (performed on silicon substrate) for obtaining the following coating on each sensor: 20\,nm, 200\,nm and 20\,nm of Ni, $^{10}$B and Ti, respectively. 
The deposition was initiated once the target cleaning has been carried out by simultaneously applying $-600$\,V of DC bias 
to the metal targets (Ni and Ti) and 440\,V of DC self-bias (250\,W of RF power) to the boron dielectric target.

For direct deposition on sensors, a special copper holder was designed to guarantee the thermal transfer and the electrical contact between the sensor and the holder during deposition, 
the latter being electrically kept grounded and thermally at the room temperature by water cooling. 
The thermal transfer was optimized using graphite based interface pad (Tgon\textsuperscript{TM} 820, from Laird Technologies) on the lateral sides. The quality of the thermal contact is essential as it will impact the quality of the sputtered layer. The CCD holder was then placed on the substrate holder and fastened in place with silver lacquer and screws.

\begin{figure}[t]
\begin{center}
\includegraphics[width=0.49\textwidth]{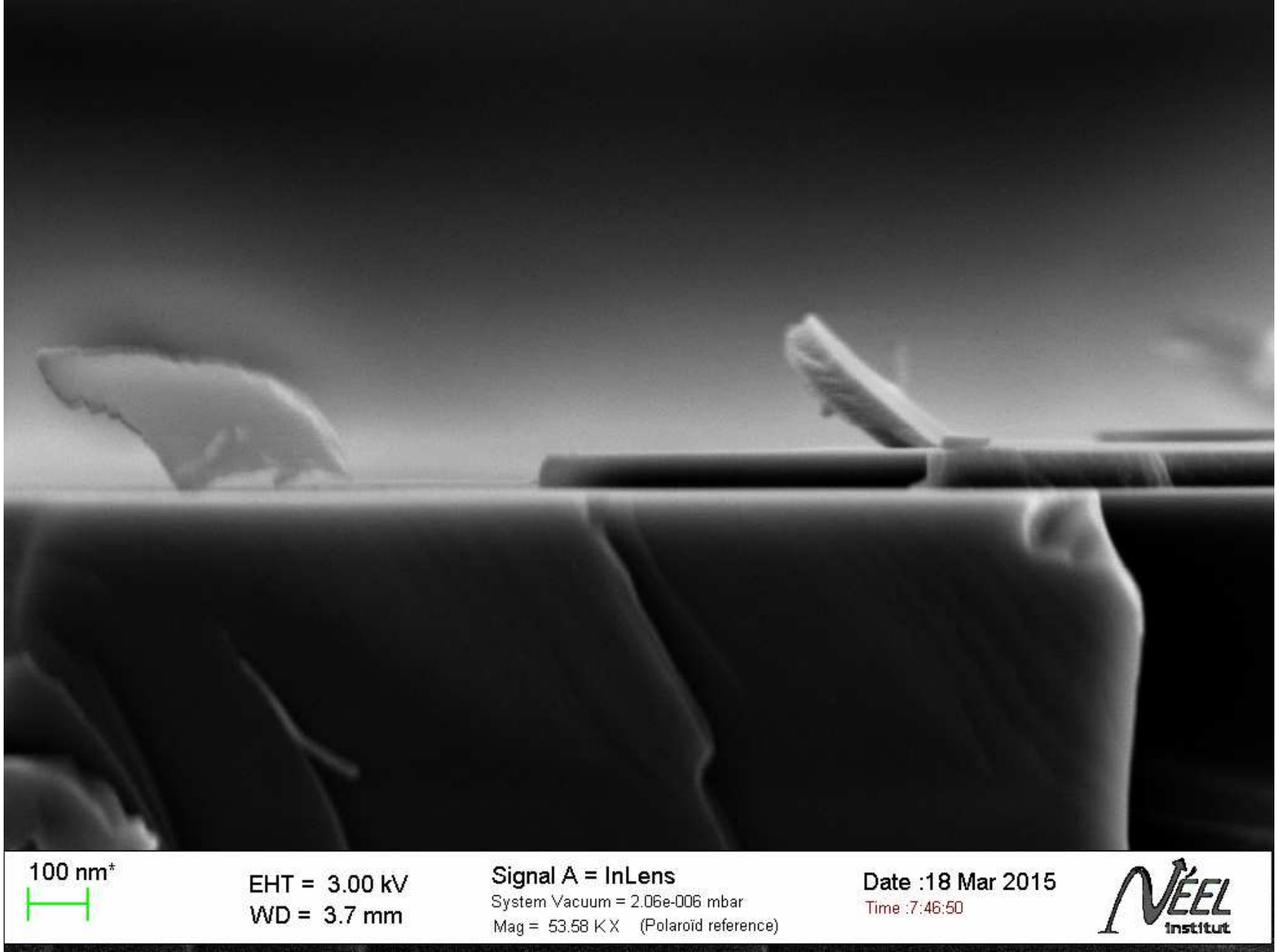}
\includegraphics[width=0.46\textwidth]{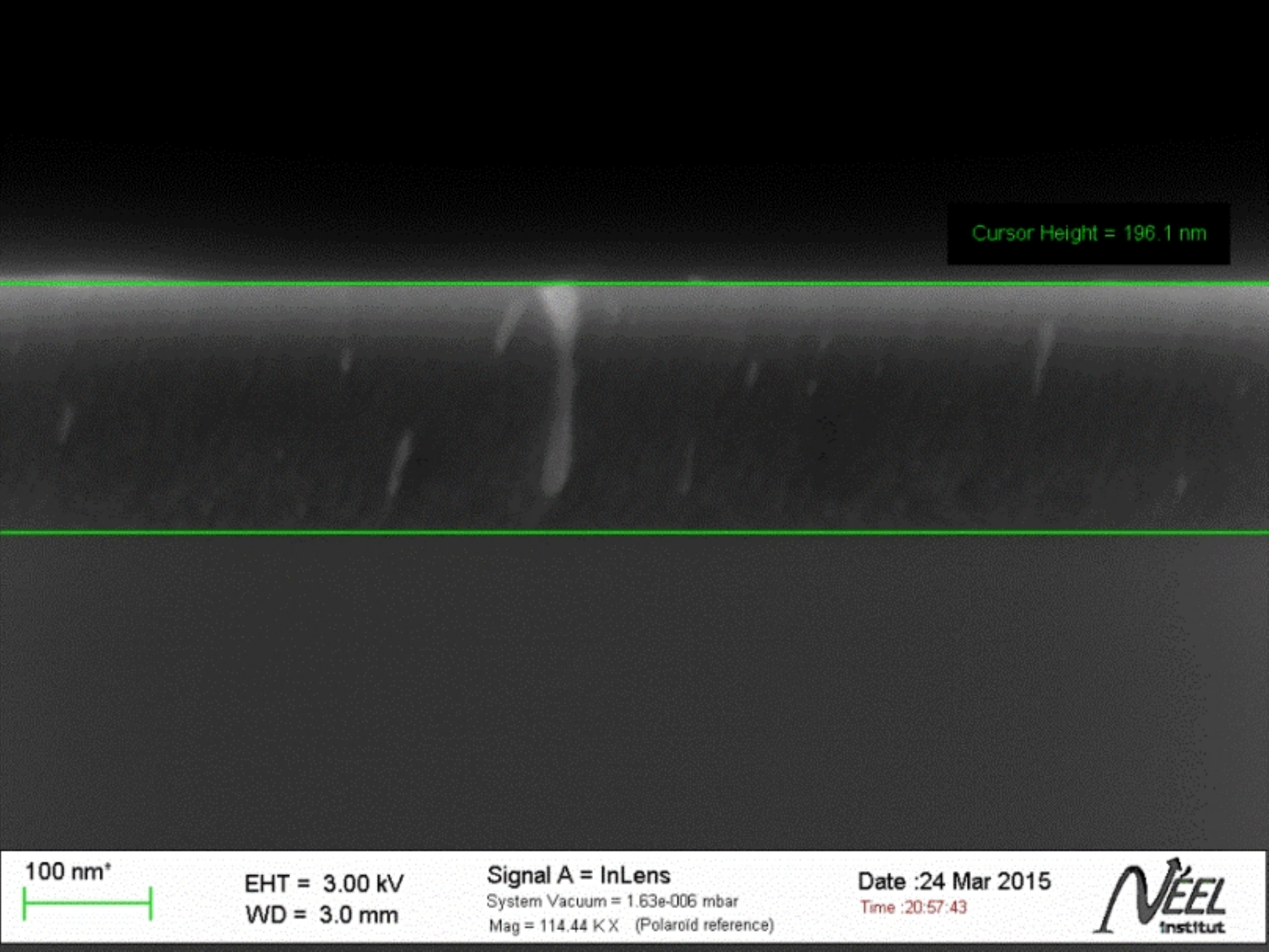}
\caption[]{Cross section view of boron films deposited on a plasma cleaned silicon substrate with 50\,W of RF polarization, for a six hours deposition process: for 10 (on the left) and 20 minutes (on the right) cleaning. In this last case, the thickness of the layer is 196\,nm.}
\label{fig-flakes}
\end{center}
\end{figure}

\begin{figure}[t]
\begin{center}
\includegraphics[width=0.45\textwidth]{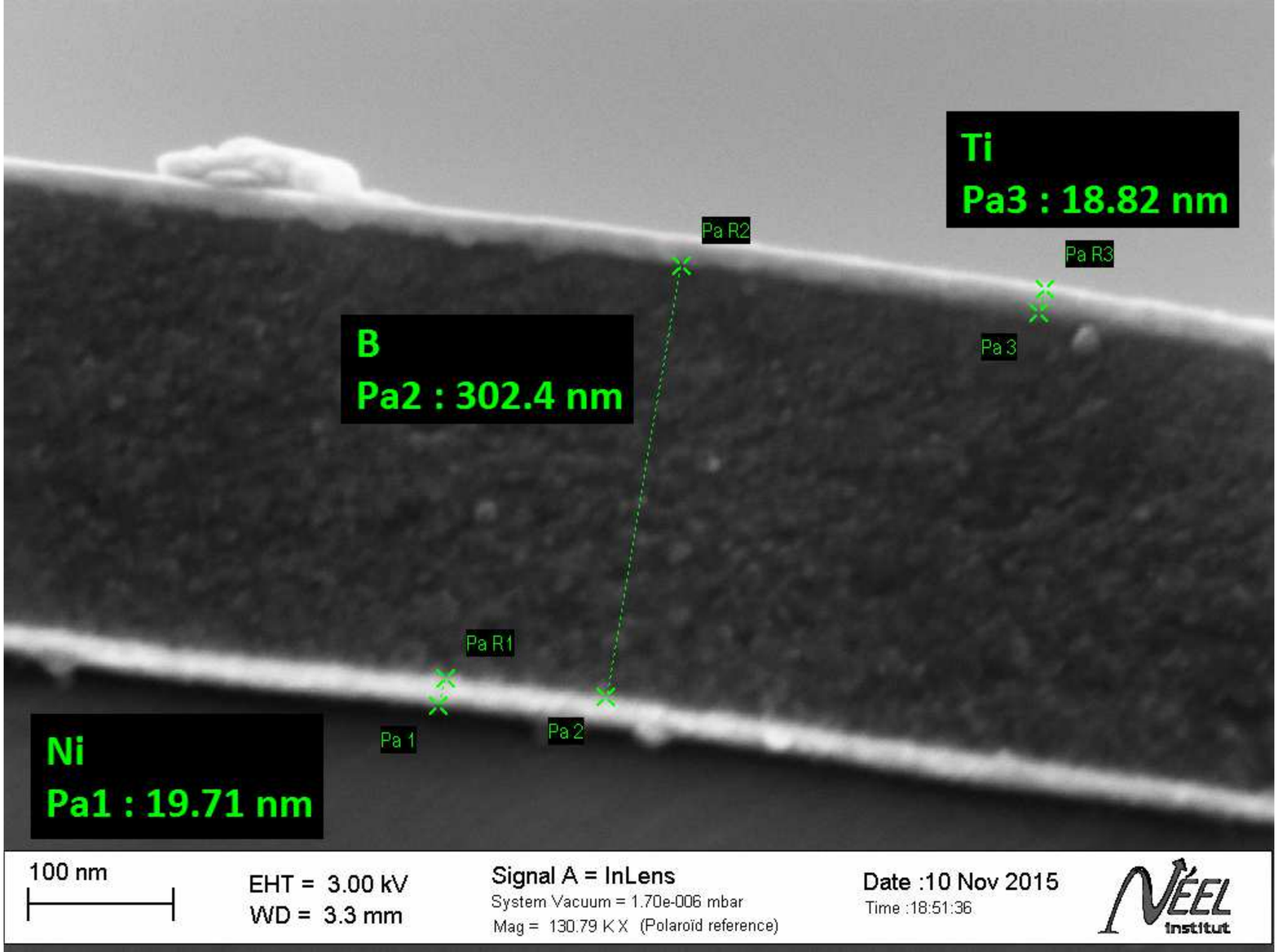}
\caption[]{A cross section of a multilayer Ti/B/Ni (20 / 300 / 20 nm) film grown onto a silicon substrate. Each layer is measured at a random point and the thickness are found to be $18.8$\,nm (Ti), $302.4$\,nm (B) and $19.7$\,nm (Ni) with a precision of 1\,nm. The uniformtity of the layer over the all sustrate is of $\pm2$\,nm for the smalle Ti and Ni layers and $\pm5$\,nm for the boron layer.}
\label{fig-multilayer}
\end{center}
\end{figure}

The microstructure of deposited films on Si substrates and also their thickness were examined by SEM (scanning electron microscope) 
using Zeiss Ultra+ Field Emission SEM (EHT voltage levels of 3\,kV). The preliminary calibration step clearly pointed out the consequence 
of the surface plasma cleaning, especially that of the substrate. 
Indeed, this leads not only to the removing of hydrocarbon impurities and/or native oxides but also to the creation of a slight surface roughness. 
In this way, the adhesion to the substrate of the growing film is greatly improved and the impact of the compressive stress strongly diminished. 
Figure~\ref{fig-flakes} compares a boron film deposited on a plasma cleaned silicon substrate for 10\,min to the one obtained with a plasma cleaning for 20\,min (argon plasma at a pressure of 0.13\,Pa and 50\,W of RF power applied to the substrate). 
For a not enough bias voltage or processing time, the grown film exhibits a compressive stress leading to the delamination in form of flakes (figure~\ref{fig-flakes}, left side). 
An appropriate plasma cleaning allows growing of a compact film with a good adhesion to the substrate surface (figure~\ref{fig-flakes}, right side). 
The same compact microstructure can be observed in figure~\ref{fig-multilayer} for a multilayer film (with a boron thickness of 300 nm) deposited on a silicon substrate. 
A good adhesion to the substrate and well delimited interfaces between the three layers can be observed.


The use of multilayer structured coatings for the stress reduction without post-deposition annealing was already reported in~\cite{Vassallo2013} 
for boron-carbon multilayer coating by RF plasma sputtering. However, the multilayer with a thickness of about 330\,nm includes six layers of the 
same B-C coating obtained in the sequential manner. The method used in this work enables to obtain the neutron reactive 
layer (of 200\,nm on sensor, with a deposition rate of about 0.75\,nm.min$^{-1}$) in a single sequence while the two other sequences are used for deposition 
on both sides of the boron film of two additional layers of different natures: Ni and Ti. The first promotes not only a increased adherence but also the reflection back into the reactive layer of the unconverted UCNs during their first crossing. 
The second protects the boron layer against oxidization thereby ensuring its chemical stability.

\begin{figure}[t]
\begin{center}
\includegraphics[width=0.45\textwidth]{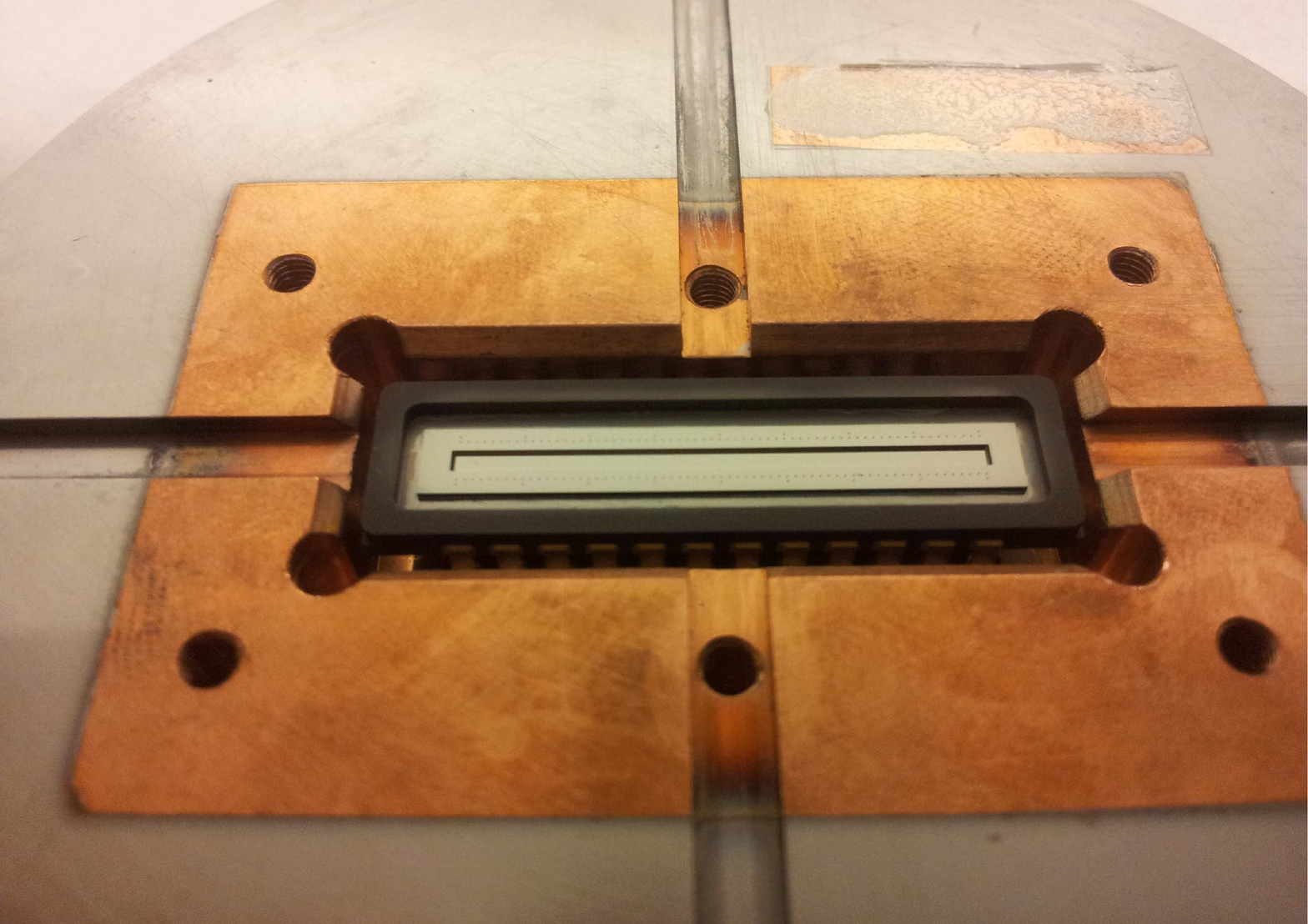}
\caption[]{A coated CCD sensor inside of the copper holder.}
\label{fig-coating}
\end{center}
\end{figure}

An example of CCD sensor coated is given in figure~\ref{fig-coating}. The functional properties of the multilayer coatings on the sensor and aging behavior were only evaluated through the CCD sensor performance as described in section~\ref{neutrons} .

\section{Tests with neutrons}
\label{neutrons}

The range in silicon of the ions produced in the neutron absorbtion on boron does not exceed a few microns with an even smaller lateral straggling. As we used a pixelized detector, the produced charge will drift within the bulk silicon and several pixels are hit. 
The position could therefore  be reconstructed using a weighted average over a cluster of pixels.
The central pixel contains between a fourth and a fifth of the total energy and 99\% is contained in a 9 by 9 pixels matrix around the center, for both lithium and helium ions. On this basis, we constructed a simple clustering algorithm. For each image, a $3\times3$ sliding window is applied to determine seed clusters whose total ADC count exceed a given threshold. Each cluster is afterwards characterized by three variables: the reconstructed horizontal and vertical positions and the sum of ADC counts in the cluster which represent the total charge in arbitrary units. A 30\,kBq $^{241}$Am $\alpha$ source attenuated through 30 to 40\,mm of air was used to calibrate the energy response. Data from NIST databases are used to estimate the alpha particle energy loss. Energy measurement in itself is not the aim for the detector, 
but it can be used to ensure that the observed signal comes from the neutron absorbtion on boron.

\begin{figure}[t]
\begin{center}
\includegraphics[width=\textwidth]{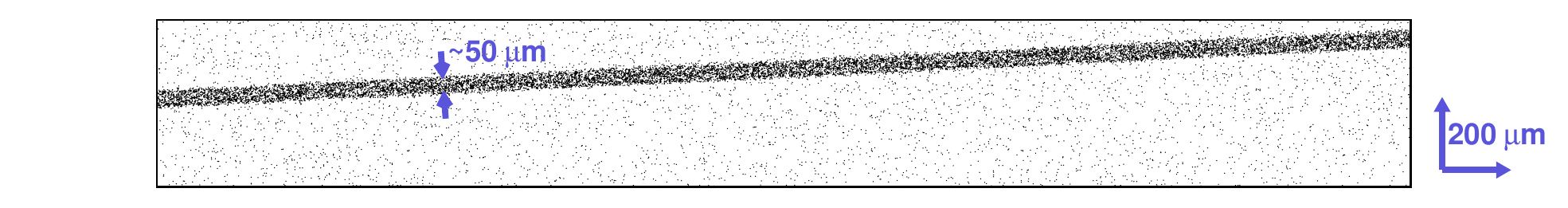}
\includegraphics[width=\textwidth]{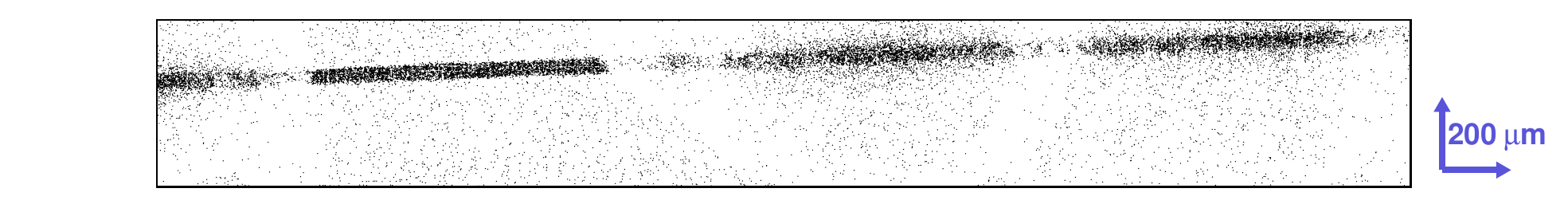}
\caption[]{Impact position of the reconstructed alpha and lithium ions from cold neutrons out of a slit. On the top picture, the increased neutron flux at the slit level is clearly seen. 
On the bottom picture, fuzzy structures appear revealing a non adhering boron layer (flakes), which was confirmed visually. The cracks on the sensor are expected to come from a faulty thermal contact in the CCD holder during the deposition.}
\label{fig-ccdslit}
\end{center}
\end{figure}

Two coated sensors were tested with neutrons at ILL several months after the deposition of the conversion layer. The first test consisted of exposing the sensors to a cold neutron beam at the PF1b beam line~\cite{Abele2006}.
This beam line produces a neutron beam with a mean wavelength of 4 to $4.5$\,\AA\,with a flux of $1.8\times10^{10}$ n.cm$^{-2}$.s$^{-1}$
Neutrons were collimated through a 50\,$\mu$m slit made by two 10\,cm long glass plates with rough surfaces to avoid the specular reflection of neutrons. 
The surface of the sensor was placed a few millimeters downstream the output of the slit. 
The slit image obtained on both sensors can be seen in figure~\ref{fig-ccdslit}. 
The first sensor had a completely adhering boron layer, whereas the other presented cracks and flakes. 
As a result a clear loss of spatial resolution and efficiency can be qualitatively observed on the second sensor. 
This measurement allows for a high flux, but is not useful to measure the efficiency as the layer is optimized for ultra cold neutrons.

\begin{figure}[t]
\begin{center}
\includegraphics[width=0.5\textwidth]{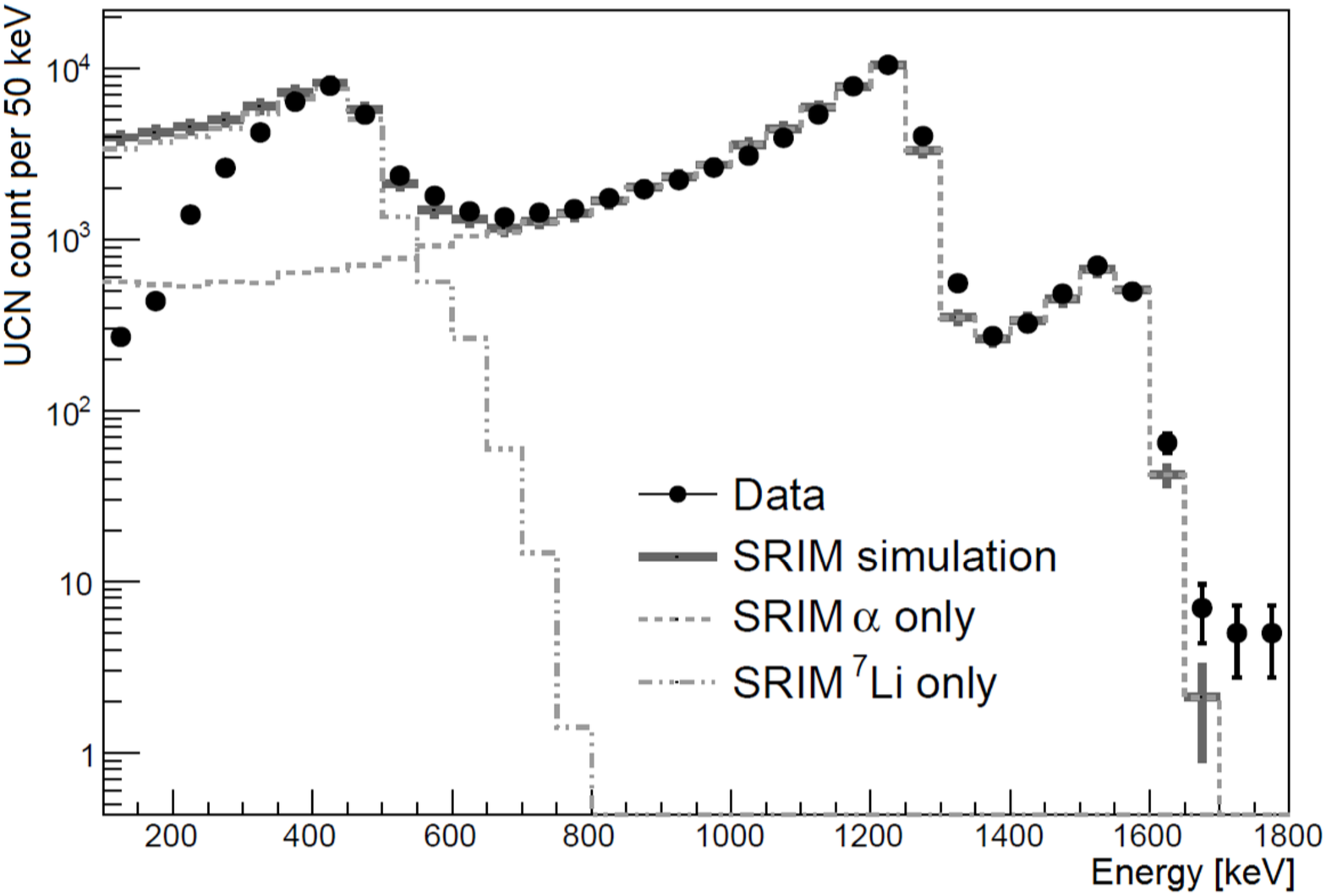}
\caption[]{An energy spectrum of the detected charged particles together with a comparison to SRIM simulations for both alpha and lithium ions. The simulation are normalized to data in the energy range $1000$ to $1300$~keV.}
\label{fig-energy}
\end{center}
\end{figure}

The coated sensor was also exposed to ultra-cold neutrons using the SUN source of the GRANIT facility at ILL.
The energy deposited in the silicon detector is shown in figure~\ref{fig-energy}. It shows the expected alpha and lithium peaks, as well as the more energetic alpha representing 6\% of the absorbtions. The left tail of the peaks comes from the varying length of boron traversed by the particle depending on the emission angle. A comparison to a SRIM simulation shows a good agreement.

The UCN flux was determined using a standard gaseous $^{3}$He UCN detector. The detector window is a 15\,$\mu$m titanium foil, which accounts for $24\pm 1\%$ neutron loss either due to absorption or quantum reflection.
This efficiency is estimated using Monte-Carlo simulation and the error is dominated by geometrical effects. 
The UCN flux on the sensor surface is therefore estimated as $14\pm 1.5$\,UCN.cm$^{-2}$.s$^{-1}$. 
The flux measured by the sensor itself is $11.41\pm0.12$\,UCN.cm$^{-2}$.s$^{-1}$.
The resulting efficiency is $82\pm9\,\%$ which is consistent with the expectation discussed in section~\ref{design}.

\section{Conclusion}
\label{conclusion}
We studied ultra-cold neutron detectors based on a $^{10}$B conversion layer coupled to a silicon detector.
Such a conversion layer needs several optimizations: thickness of the boron layer including both the conversion efficiency and the charged particles energy loss in the layer, effect of the passivisation layer, reflective gain from the nickel layer. We showed that a multilayer composed of a 20\,nm titanium protection layer on top a 200\,nm enriched boron-10 and 20\,nm nickel should provide a total conversion efficiency from 83\% to 89\% over a large range of neutron velocities. To this end, a reliable process for boron deposition based on cold plasma sputtering has been developed. The multi target configuration of the plasma reactor allows for the production of the multilayer in one single process. Such a multilayer was produced on the surface of a CCD sensor. Neutrons measurement were performed at PF1b and GRANIT instrument at ILL providing respectively cold and ultracold neutrons to test the performances of the sensor. The total efficiency is found to be $82\pm9\,\%$ which is in agreement with our expectations. We also showed that neutrons can be used to check the uniformity and adherence of the layer on a position sensitive detector. Nevertheless, the use of this sensor as a position sensitive detector will now require more detailed study of its spatial resolution.

\bibliographystyle{JHEP}
\bibliography{boronCoatedSi} 
\end{document}